\documentclass[sigconf]{acmart}
\renewcommand\footnotetextcopyrightpermission[1]{} 
\usepackage{dcolumn}
\usepackage{multirow}
\usepackage{booktabs}
\usepackage{dcolumn} 
\newcolumntype{d}[1]{D..{#1}}
\usepackage{bm}

\usepackage{xltabular}

\begin{document}

\title{Quantifying the topic disparity of scientific articles}





\author{Munjung Kim}
\affiliation{%
  \institution{\textit{Department of Physics \\ Pohang University of Science and Technology}}
  \streetaddress{77 Cheongam-ro}
  \city{Pohang-si}
  \country{Republic of Korea}}

\author{Jisung Yoon}
\affiliation{%
  \institution{\textit{Department of Industrial and Management Engineering\\ Pohang University of Science and Technology}}
  \streetaddress{77 Cheongam-ro}
  \city{Pohang-si}
  \country{Republic of Korea}}

\author{Woo-Sung Jung}
\affiliation{%
  \institution{\textit{Department of Physics\\Department of Industrial and Management Engineering\\Pohang University of Science and Technology}}
  \city{Pohang-si}
  \country{Republic of Korea}}
\email{wsjung@postech.ac.kr}

\author{Hyunuk Kim}
\affiliation{%
  \institution{\textit{Department of Administrative Sciences\\Metropolitan College, Boston University}}
  \city{Boston}
  \state{MA}
  \country{USA}}
\email{uk@bu.edu}

\renewcommand{\shortauthors}{Kim et al.}

\begin{abstract}
Citation count is a popular index for assessing scientific papers. However, it depends on not only the quality of a paper but also various factors, such as conventionality, team size, and gender. Here, we examine the extent to which the conventionality of a paper is related to its citation percentile in a discipline by using our measure, topic disparity. The topic disparity is the cosine distance between a paper and its discipline on a neural embedding space. Using this measure, we show that the topic disparity is negatively associated with the citation percentile in many disciplines, even after controlling team size and the genders of the first and last authors. This result indicates that less conventional research tends to receive fewer citations than conventional research. Our proposed method can be used to complement the raw citation counts and to recommend papers at the periphery of a discipline because of their less conventional topics. 

\end{abstract}

\ccsdesc[500]{Information systems~Document representation}
\ccsdesc[500]{Applied computing~Law, social and behavioral sciences}

\keywords{Neural embedding techniques, BERT, Microsoft Academic Graph}

\maketitle

\section{Introduction}

Citations have been used to evaluate articles, individual researchers, and organizations~\cite{moed2006citation,cole2000short,fersht2009most}. The assessments are based on the perspective that high quality research receives more citations than low quality research~\cite{bornmann2008citation}. However, citations also vary by team size, gender, and conventionality. Large team is likely to receive more citations compared to small team~\cite{lariviere2015team,wuchty2007increasing,wu2019large}, and female-authored papers tend to be less cited than male-authored papers~\cite{dworkin2020extent,teich2021citation,fulvio2021gender}. In addition, papers with highly conventional pairs tend to receive more citations than papers with unconventional pairs~\cite{uzzi2013atypical}. Papers with new pairs have greater chances of being highly cited but also have higher variances of citations~\cite{wang2017bias}, suggesting that novel research would be both risky and impactful. 

Among the factors above, we focus on conventionality and suggest an alternative approach to quantifying conventionality by leveraging a neural embedding method that represents scientific texts as vectors~\cite{cohan2020specter}. Our measure is called \textit{topic disparity} and based on actual texts rather than journal pairs. The topic disparity is defined as the cosine distance between a paper and its discipline on a vector space. Hence, the smaller the topic disparity is, the more conventional a paper is. We show that the topic disparity is negatively correlated with the citation percentile in many disciplines, even team size and the genders of the first and last authors are considered. Our analysis not only confirms that conventional research tends to receive more citations than less conventional research~\cite{uzzi2013atypical} but also highlights differences in citing patterns by discipline.

\begin{figure*}[!]
  \includegraphics[width=0.8\textwidth]{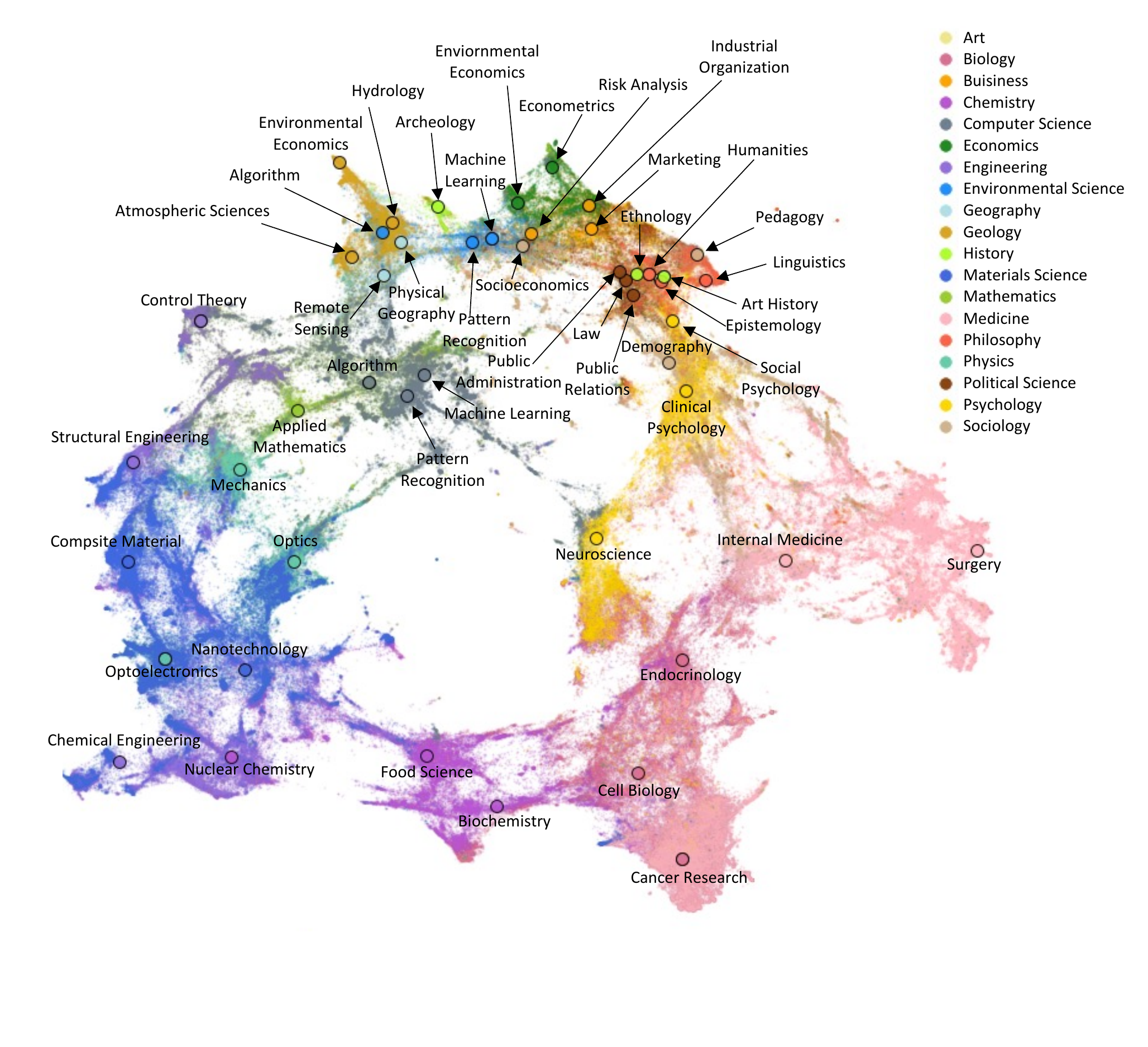}
  \caption{A UMAP~\cite{mcinnes2018umap} projection of 49 discipline vectors and their 746,654 papers. Each small and large points correspond to a paper and a selected discipline, respectively.
  }
  \label{fig:UMAP1}
\end{figure*}

\section{Data and Methods}\label{sec2}
\subsection{Microsoft Academic Graph}\label{sec:data}

We retrieve journal articles published in 2019 from Microsoft Academic Graph (MAG; accessed on November 2, 2021). MAG is discontinued on December 31, 2021 and migrated to OpenAlex afterwards. MAG provides hierarchical discipline information of each article~\cite{sinha2015overview}, namely ``Field of Study (FoS)'' ranging from Level 0 (Highest) to Level 5 (Lowest). 
Level 0 and Level 1 codes are named as fields and disciplines hereafter. A paper can belong to multiple fields and disciplines.
There are 19 fields: Art, Biology, Business, Chemistry, Computer Science, Economics, Engineering, Environmental Science, Geography, Geology, History, Materials Science, Mathematics, Medicine, Philosophy, Physics, Political Science, Psychology, and Sociology. 

\subsection{Topic disparity}\label{sec:topic_disparity}

We convert titles and abstracts of the papers into real-valued vectors by using SPECTER~\cite{cohan2020specter}, a transformer-based machine learning model trained on scientific citations. We define the discipline vector $V_i$ as the mean vector of papers that belong to discipline $i$ as follows, 


\begin{equation}
   V_{i} = \frac{\sum_{j} v_{i,j}}{N_{i}},
\end{equation}

where $N_i$ is the total number of papers in discipline $i$ and $v_{i,j}$ is the embedding vector of paper $j$ in discipline $i$. We assumed that $V_i$ is the overall theme of discipline $i$.

The vector representations of papers and disciplines allow us to calculate $D$ the topic disparity of a paper from its discipline. We define $D_{i,j}$ as the cosine distance between the vector of paper $j$ and the vector of discipline $i$ to which $j$ belongs. We limit our analysis to top three disciplines for each field with respect to the number of papers. ``Algorithm'', ``Art History'', ``Cancer Research'', ``Control Theory'', ``Environmental Planning'',  ``Humanities'', ``Industrial Organization'', and ``Optoelectronics'' belong to two disciplines, so 49 disciplines under 19 fields are used in our analysis.


\begin{figure*}[!]
  \includegraphics[width=1\textwidth]{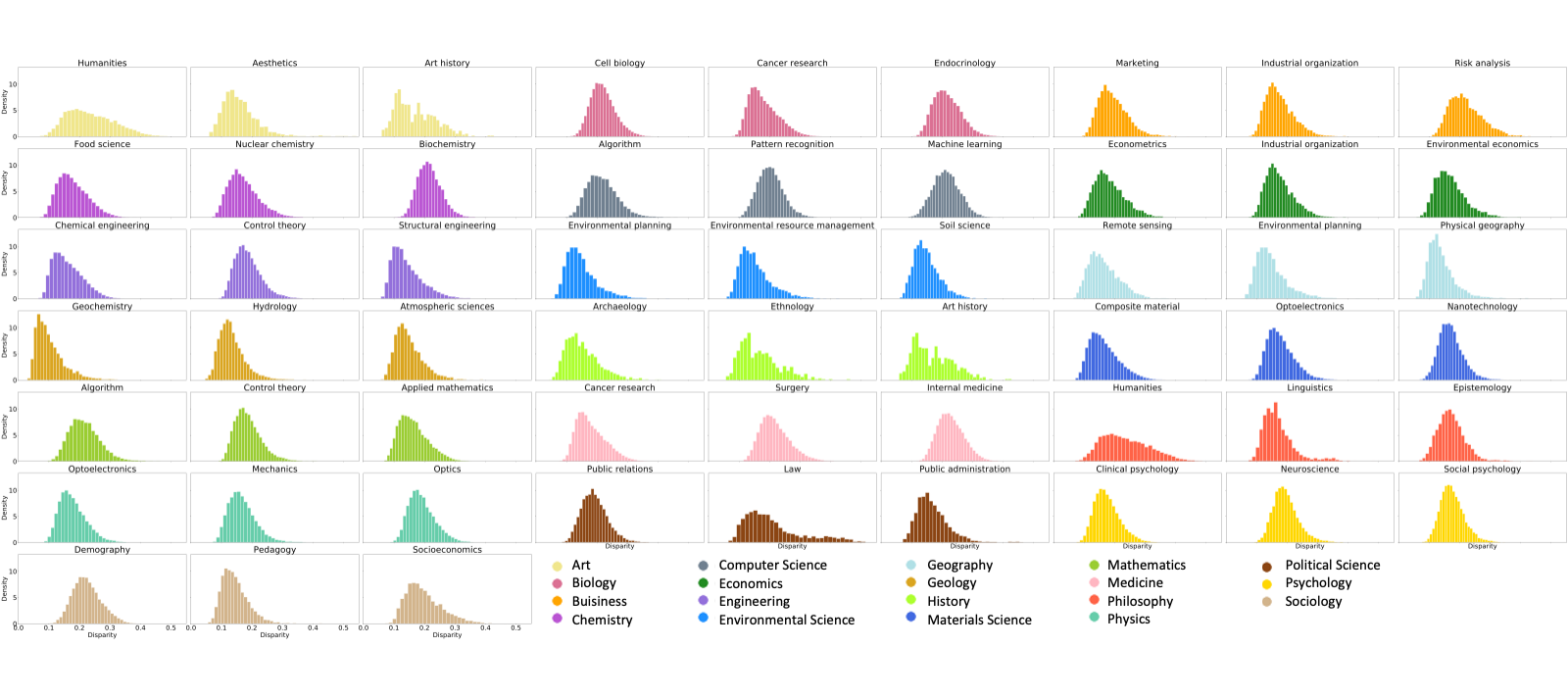}
  \caption{The distributions for the topic disparity values of the 49 disciplines. The disciplines which belong to two fields are shown twice in different colors.}
  \label{fig:Distribution}
\end{figure*}

\subsection{Genders of the first and last authors}\label{sec:gender}

We use Genderize (\href{https://genderize.io/}{https://genderize.io/}) to infer the genders of the first and last authors of the papers in the selected 49 disciplines. For each given name, we assign either female or male if the probability returned from Genderize is higher than 0.7. If not, we use the Wiki-Gendersort algorithm~\cite{berube2020wiki} to fill missing genders as much as possible. Initials are removed from given names in order to reduce noises. 


Papers are then categorized into FF, FM, MF, or MM, depending on the genders of the first and last authors. F and M stand for female and male, respectively. For instance, if a paper is written by a female first author and a female last author, this paper is classified as FF. Although the proportions of these categories vary by discipline, in total, FF and MF have smaller numbers of papers than FM and MM (13.5\% FF, 25.1\% FM, 15.0\% MF, 46.3\% MM).

\subsection{Linear regression}

To examine the impact of topic disparity on citations, we build linear regression models with team size and the genders of the first and last authors by discipline as follows,

\begin{equation}
    C_{i,j} = \beta_{0i} +\beta_{1i} n_{i,j}+ \beta_{2i} \text{D}_{i,j}+\beta_{3i} \text{FF}_{i,j} + \beta_{4i} \text{FM}_{i,j}+ \beta_{5i} \text{MF}_{i,j} + e_{i,j},\label{eq:gender}
\end{equation}

where $C_{i,j}$ is the percentile of the citation counts of paper $j$ in discipline $i$, $n_{i,j}$ is the number of authors, and $\text{D}_{i,j}$ is the topic disparity of paper $j$ in discipline $i$. $\text{FF}_{i,j}$, $\text{MF}_{i,j}$, and $\text{FM}_{i,j}$ are three dummy variables indicating the genders of the first and last authors. For instance, \text{FF} is 1 if paper $j$ in discipline $i$ is written by female first and last authors. We exclude papers of which genders are not inferred for any author or the number of authors is larger than the 99th percentile in a discipline to minimize the impact of missing values and outliers. As a result, 746,654 papers from the 49 disciplines are used in the regressions.

\section{Results}\label{sec3}

The paper and discipline vectors are projected onto a two dimensional space by the UMAP algorithm~\cite{mcinnes2018umap} (Figure~\ref{fig:UMAP1}). Large and small points correspond to the 49 disciplines and their 746,654 papers, respectively. Overall, the projection is consistent with existing maps of science~\cite{rafols2010diversity,borner2012design,murdock2017multi}.

For all disciplines, the distribution of topic disparities has a peak in the middle of the value range and is right-skewed (Figure~\ref{fig:Distribution}), implying most papers combine the overall theme of a discipline with less conventional components, while there are a small portion of papers pursuing genuinely novel topics.

From the regressions, we find that the coefficient of the topic disparity is negative for 35 out of 49 disciplines ($p<0.05$, Table~\ref{tab:regression}). It suggests that papers of high disparity values tend to receive less citations than papers of low disparity values in many disciplines. The relationship between the disparity and the citation percentile is significantly positive in six disciplines ($p<0.05$, Table~\ref{tab:regression}): Aesthetics, Art history, Econometrics, Humanities, Laws, and Linguistics. 


Also, we find that team size is positively associated with the citation percentile in 48 out of 49 disciplines ($ p < 0.01$; Table~\ref{tab:regression}), indicating large teams tend to receive more citations than small teams. The variable FF is significant in 28 out of 49 disciplines ($p<0.05$; Table~\ref{tab:regression}) and their coefficients are all negative, suggesting papers written by female first and last authors tend to have less citations than other papers in these disciplines. In the case of FM, significant associations are found in 18 disciplines ($p<0.05$; Table~\ref{tab:regression}). Among them, the estimated coefficients in 16 disciplines were negative. The coefficients of MF are significant in 14 disciplines ($p<0.05$; Table.~\ref{tab:regression}) and all of them are negative. 

\section{Conclusion}

Here, we present a method for measuring the topic disparity of a paper by calculating the cosine distance between the paper and its discipline on a vector space. By applying this method, we examined the relationship between the topic disparity and the citation percentile while considering team size and gender.

Our result shows that there is a negative relationship between the citation percentile and the topic disparity in 35 out of 49 disciplines. Therefore, research papers focusing on different topics from the main research theme of a given discipline tend to receive fewer citations than papers on conventional topics. A potential explanation is that areas studying less conventional topics are relatively small, so authors may receive fewer citations~\cite{moed1985use,king1987review}.

The relationship between the citation percentile and team size is significantly positive in all disciplines except ``Aestetics''. This matches well with what have been revealed from previous studies~\cite{lariviere2015team,wuchty2007increasing}. Also, the research conducted by female first and last author tends to receive significantly less citation than other types of research in 28 out of 49 disciplines. This result is consistent with existing studies that show female-authored research receives less citations than male-authored research~\cite{caplar2017quantitative,lariviere2013bibliometrics,dworkin2020extent,teich2021citation}. 

Our approach can be extended to investigate the relationships between the topic disparity and other attributes in science such as the accessibility of research papers, the nationality and ethnicity of authors, and the journals where articles are published. We also expect that the topic disparity identifies marginalized papers and researchers developing novel perspectives. 

\clearpage
\onecolumn

\begin{table*}[!]
\caption{Regression results for the selected 47 disciplines}
\label{tab:regression} 
{\footnotesize
\begin{xltabular}{17 cm}{p{5 cm} | c c c c c c}
\multicolumn{7}{@{}l}{\em}\\
\multicolumn{1}{l}{Discipline (Number of Observations )} &  Team Size & Topic Disparity & FF& FM& MF &$R^2$\\ \hline\hline

Aesthetics (1,075)& $0.015(0.011)$& 
$1.111^{***}(0.144)$ &
$0.008(0.022)$ &
$-0.009(0.023)$ &
$-0.012(0.024)$ &
$0.057$ \\\hline

Algorithm (18,788)& $0.020^{***}(0.001)$& 
$-0.289^{***}(0.041)$ &
$-0.053^{***}(0.008)$ &
$-0.025^{***}(0.005)$ &
$-0.023^{***}(0.006)$ &
$0.016$ \\\hline

Applied Mathematics (9,943)& $0.034^{***}(0.003)$ & $-0.301^{***}(0.060)$ &
$-0.071^{***}(0.012)$&
$-0.043(0.008)$&
$-0.030(0.008)$&
$0.021$   \\\hline

Archaeology (2,043)& $0.027^{***}(0.002)$ & $-0.288^{***}(0.108)$ &
$0.003(0.018)$&
$0.001(0.015)$&
$-0.001(0.017)$&
$0.076$   \\\hline

Art History (516)& $0.020^{***}(0.005)$ & $0.660^{***}(0.163)$ &
$0.025(0.027)$&
$0.005(0.028)$&
$0.038(0.028)$&
$0.066$   \\\hline

Atmospheric Science (4,719)& $0.018^{***}(0.001)$& 
$-0.406^{***}(0.089)$ &
$-0.034^{*}(0.014)$ &
$-0.008(0.010)$ &
$-0.012(0.012)$ &
$0.059$ \\\hline




Biochemistry (14,863)& 
$0.012^{***}(0.001)$ & 
$-0.309^{***}(0.060)$ &
$-0.021^{**}(0.007)$&
$-0.012^{*}(0.006)$&
$-0.011(0.007)$&
$0.017$ \\\hline


Cancer Research	 (44,362) 
&$0.008^{***}(0.000)$ & 
$-0.369^{***}(0.028)$&
$-0.002(0.004)$&
$0.009^{**}(0.003)$&
$0.007(0.004)$ & $0.021$ \\\hline


Cell biology (45,093) & $0.015^{***}(0.000)$& $ -0.348^{***}(0.033)$ &
$-0.015^{***}(0.004)$ &
$-0.010^{**}(0.003)$ &
$0.002(0.004)$&
$0.042$  \\\hline

Chemical Engineering (52,929) & $0.023^{***} (0.001) $ & 
$-0.969^{***} (0.025)$ & $-0.017^{***}(0.004)$ & $-0.006^{*}(0.003)$ & $-0.008^{*}(0.004)$ &$0.071$ \\\hline

Clinical Psychology (18,263) & $0.016^{***} (0.001) $ & $-0.830^{***} (0.048)$ & $-0.016^{***}(0.006)$ & $-0.004(0.006)$ & $-0.010(0.007)$ &$0.039$ \\\hline




Composite Material (36,788)& $0.024^{***}(0.001)$ &
$-0.540^{***}(0.030)$&
$-0.019^{**}(0.006)$&
$-0.004(0.004)$&
$-0.008(0.004)$& $0.036$ \\\hline




Control theory (18,822)& 
$0.021^{***}(0.002)$ & 
$-0.668^{***}(0.053)$ & 
$-0.030^{**}(0.010)$&
$-0.006(0.006)$&
$-0.014^{*}(0.007)$&
$0.020$ \\\hline

Demography (17,162)& 
$0.015^{***}(0.001)$ & 
$-0.698^{***}(0.046)$ & 
$-0.018^{**}(0.006)$&
$-0.010(0.006)$&
$-0.007(0.006)$&
$0.044$  \\\hline



Econometrics (6,176) & 
$0.037^{***}(0.004)$& 
$0.155^{*}(0.072)$ &
$-0.085^{***}(0.014)$&
$-0.037^{***}(0.010)$&
$-0.022^{*}(0.010)$&
$0.024$ \\\hline

Endocrinology (26,897)&
$0.018^{***}(0.001)$ &
$-0.679^{***}(0.037)$&
$0.002(0.005)$&
$0.008(0.004)$&
$-0.008(0.005)$&
$0.063$ \\\hline

Environmental Economics (26,897	)& 
$0.025^{***}(0.003)$&
$-0.977^{***}(0.083)$ & 
$-0.040^{***}(0.014)$ &
$-0.024^{*}(0.010)$&
$-0.021(0.012)$&
$0.052$
\\\hline



Environmental Planning (3,891)& $0.025^{***}(0.002)$ & 
$-0.298^{***}(0.083)$ & 
$-0.015(0.012)$&
$-0.001(0.011)$&
$-0.012^{**}(0.012)$&
$0.068$  \\\hline

Environmental Resource Management (2,856)& 
$0.018^{***}(0.001)$ & 
$-0.145(0.097)$ & 
$-0.057^{***}(0.016)$&
$-0.011(0.013)$&
$-0.018(0.015)$&
$0.065$  \\\hline

Epistemology (2,548)& 
$0.027^{***}(0.005)$ & 
$0.098(0.125)$ & 
$0.001(0.015)$&
$-0.005(0.015)$&
$-0.029(0.016)$&
$0.011$  \\\hline

Ethnology	 (580)& 
$0.024^{***}(0.004)$ & 
$0.219(0.174)$ & 
$0.020(0.031)$&
$0.003(0.028)$&
$-0.010(0.033)$&
$0.062$  \\\hline


Food Science (18,822)& 
$0.023^{***}(0.001)$ &
$-0.376^{***}(0.041)$&
$-0.024^{***}(0.006)$&
$-0.009(0.005)$&
$-0.010(0.006)$ & $0.036$ \\\hline


Geochemistry (4,864)&
$0.024^{***}(0.002)$&
$-0.426^{***}(0.085)$&
$-0.035^{*}(0.015)$&
$-0.027^{***}(0.010)$&
$-0.010(0.011)$&
$0.049$  \\\hline




Humanities (8,042) &
$0.019^{***}(0.002)$ & 
$0.479^{***}(0.038)$ &
$-0.003(0.007)$&
$0.003(0.007)$&
$-0.009(0.008)$&
$0.057$ \\\hline

Hydrology (4,850) &
$0.024^{***}(0.002)$ & 
$-0.555^{***}(0.093)$ &
$-0.013(0.014)$&
$0.005(0.010)$&
$0.001(0.012)$&
$0.040$ \\\hline

    
Industrial Organization	(4,553)
&$0.031^{***}(0.004)$&
$-0.480^{***}(0.090)$&
$-0.061^{***}(0.013)$
&$-0.021(0.011)$ &$-0.008(0.011)$ & $0.022$ \\\hline


Internal Medicine (38,510)& 
$0.014^{***}(0.000)$ & 
$-0.594^{***}(0.032)$ & 
$-0.0221^{***}(0.004)$&
$-0.005(0.003)$&
$-0.014^{**}(0.004)$&
$0.060$ \\\hline

Law (2,411)& 
$0.016^{***}(0.004)$ & 
$0.953^{***}(0.058)$ & 
$-0.031^{*}(0.015)$&
$-0.006(0.014)$&
$-0.011(0.015)$&
$0.126$ \\\hline

Linguistics (2,731)& 
$0.033^{***}(0.005)$ & 
$0.615^{***}(0.102)$ & 
$-0.002(0.014)$&
$0.015(0.015)$&
$0.015(0.015)$&
$0.039$ \\\hline

Machine Learning (8,091)& 
$0.022^{***}(0.001)$ &
$-0.541^{***}(0.070)$&
$-0.079^{***}(0.012)$&
$-0.030^{***}(0.008)$&
$-0.014(0.009)$ & 
$0.041$ \\\hline

Marketing (7,350)& 
$0.039^{***}(0.003)$ & 
$-0.605^{***}(0.068)$ & 
$-0.017(0.009)$&
$-0.010(0.008)$&
$-0.004(0.009)$&
$0.033$ \\\hline



Mechanics (19,014)& 
$0.020^{***}(0.001)$ &
$-0.604^{***}(0.045)$&
$-0.055^{***}(0.010)$&
$-0.020^{***}(0.006)$&
$-0.020^{**}(0.006)$ & 
$0.021$ \\\hline


    
Nanotechnology (11,046)& 
$0.007^{***}(0.001)$ & $-0.726^{***}(0.068)$ &
$-0.032^{***}(0.009)$&
$-0.010(0.007)$ &
$-0.017^{*}(0.008)$&
$0.016$ \\ \hline
    
Neuroscience (14,190)& $0.010^{***}(0.001)$ & $-0.710^{***}(0.056)$ &
$-0.012(0.007)$ &
$-0.015^{**}(0.006)$&
$-0.007(0.007)$&
$0.029$ \\\hline
    
Nuclear Chemistry (15,921)& $0.022^{***}(0.001)$ & $-1.257^{***}(0.044)$ &
$-0.017^{*}(0.007)$ &
$-0.001(0.005)$&
$-0.014^{*}(0.007)$&
$0.076$  \\\hline


Optics (14,373)	 & $0.014^{***} (0.001)$& $-0.397^{***} (0.054)$ &$-0.025^{*}(0.010)$&$-0.018^{**}(0.006)$ & $-0.008(0.007)$ & $0.019$  \\\hline

Optoelectronics (21,532)& 
$0.021^{***}(0.001)	$ & 
$-0.675^{***}(0.043)$ &
$0.006(0.008)$&
$-0.004(0.005)$&
$0.009(0.006)$&
$0.063$ \\\hline


Pattern Recognition (13,498)& $0.027^{***} (0.001)$& $ 0.069 (0.059)$ &$-0.047^{***}(0.009)$&$-0.040^{***}(0.006)$& $-0.014^{*}(0.007)$& $0.034$\\\hline

      

Pedagogy (3,717)& 
$0.019^{***}(0.004)$ & 
$-0.802(0.045)$ &
$-0.010(0.006)$&
$0.001(0.005)$&
$-0.009(0.006)$&
$0.037$ \\\hline

Physical Geography (3,483)& 
$0.019^{***}(0.002)$ & 
$-0.638^{***}(0.100)$ &
$-0.051^{***}(0.017)$&
$-0.023^{***}(0.011)$&
$-0.023^{***}(0.013)$&
$0.046$ \\\hline



Public Administration (2,332)& 
$0.019^{***}(0.005)$ &
$-0.150(0.109)$&
$-0.019(0.016)$ & 
$-0.010(0.015)$&
$-0.007(0.016)$&
$0.006$ \\\hline
     
Public Relations (10,164)& 
$0.019^{***}(0.002)$ &
$-0.363^{**}(0.064)$&
$-0.023^{**}(0.007)$ & 
$-0.024^{**}(0.007)$&
$-0.026^{**}(0.008)$&
$0.019$ \\\hline


Remote Sensing (4,476)& 
$0.017^{***}(0.002)$ & 
$-1.071^{***}(0.080)$ & 
$-0.026(0.015)$&
$-0.009(0.010)$&
$-0.013(0.012)$&
$0.070$ \\\hline

Risk Analysis (3,826)& 
$0.020^{***}(0.002)$ & 
$0.067(0.084)$ & 
$-0.059^{***}(0.015)$&
$-0.005(0.011)$&
$-0.037^{***}(0.013)$&
$0.034$ \\\hline

Social Psychology (10,686)& 
$0.028^{***}(0.002)$ & 
$-0.468^{***}(0.068)$&
$-0.048^{***}(0.007)$&
$-0.018^{*}(0.007)$&
$-0.037^{***}(0.008)$ & 
$0.025$ \\\hline

Socioeconomics (3,087)&
$0.028^{***}(0.002)$ & 
$-0.044(0.084)$&
$0.000(0.014)$&
$0.018(0.012)$&
$0.018(0.014)$ & 
$0.054$ \\\hline

Soil Science (2,531)& 
$0.023^{***}(0.003)$ & 
$-0.534^{***}(0.131)$ & 
$-0.014(0.022)$&
$-0.024(0.014)$&
$0.000(0.016)$&
$0.031$ \\\hline

Structural Engineering	(8,015)& 
$0.023^{***}(0.002)$ & 
$-0.567^{***}(0.061)$&
$0.017(0.015)$&
$0.009(0.009)$&
$-0.006(0.010)$ & 
$0.020$ \\\hline

Surgery	 (39,402)& 
$0.022^{***}(0.001)$ &
$-0.402^{***}(0.028)$&
$-0.010(0.006)$&
$0.011^{**}(0.003)$&
$-0.028^{***}(0.004)$&
$0.059$ \\\hline
    

\multicolumn{1}{l}{} &   &  & & &  &\\ 
\multicolumn{1}{l}{$^{*} p<0.05$; $^{**} p<0.01$; $^{***} p<0.001$}   &  & & &  &\\    





\end{xltabular}

}
\end{table*}

\clearpage
\twocolumn

\bibliographystyle{unsrt}
\bibliography{main}

\end{document}